# Magnetic Resonance Fingerprinting Reconstruction Using Recurrent Neural Networks

Elisabeth HOPPE[a,1], Florian THAMM[a,1], Gregor KÖRZDÖRFER[b], Christopher SYBEN[a], Franziska SCHIRRMACHER[a], Mathias NITTKA[b], Josef PFEUFFER[b], Heiko MEYER[b] and Andreas MAIER[a]

[a] *Pattern Recognition Lab, Department of Computer Science, Friedrich-Alexander-Universität Erlangen-Nürnberg, Erlangen, Germany*
[b] *Siemens Healthcare, Application Development, Erlangen, Germany*

**Abstract.** Magnetic Resonance Fingerprinting (MRF) is an imaging technique acquiring unique time signals for different tissues. Although the acquisition is highly accelerated, the reconstruction time remains a problem, as the state-of-the-art template matching compares every signal with a set of possible signals. To overcome this limitation, deep learning based approaches, e.g. Convolutional Neural Networks (CNNs) have been proposed. In this work, we investigate the applicability of Recurrent Neural Networks (RNNs) for this reconstruction problem, as the signals are correlated in time. Compared to previous methods based on CNNs, RNN models yield significantly improved results using in-vivo data.

**Keywords.** Magnetic Resonance Fingerprinting, Magnetic Resonance Fingerprinting Reconstruction, Recurrent Neural Networks, Artificial Neural Networks

## 1. Introduction

Magnetic Resonance (MR) Imaging techniques mainly provide qualitative images, that mostly are $T_1$ or $T_2$ weighted. Thus, only qualitative contrast between different tissues is shown, without a real quantitative information which can be an important biomarker. In contrast, Magnetic Resonance Fingerprinting (MRF) is a recently proposed technique to generate quantitative parameter maps of physical relaxation times $T_1$ and $T_2$. During its acquisition, imaging parameters such as Flip Angle (FA) or Repetition Time (TR) are varied to create unique time signals (fingerprints) per voxel for the different relaxation times of the underlying tissues [1]. Although spiral readouts and high undersampling [2] enable fast acquisition, the state-of-the-art reconstruction suffers from large computational effort. Acquired fingerprints are compared with a dictionary of pre-simulated fingerprints [1], which is subject to the following limitations: The dictionary only contains a limited number of possible fingerprints due to storage and computational limitations. Reconstruction time is dependent on the number of dictionary entries, and furthermore, it can only result in discrete parameters contained in the dictionary. If the true fingerprint's parameters are not included in the dictionary, the reconstructed result will be erroneous [3]. In order to overcome this limitation and to be able to provide matching results without the incorporation of a large dictionary, various approaches using Deep Learning (DL) were proposed. The main advantages of DL based approaches are two-fold: (1) The short computation time, which is independent of the number of the signals contained in the dictionary, and (2) the more efficient representation of the dictionary, while providing continuous results. The methods mainly focus on applying Fully Connected Neural Networks (FNNs) [4], Convolutional Neural Networks (CNNs) [5,6] or other architectures, e.g. [7,8]. Furthermore, the influence of magnitude or complex-valued signals as input to the networks is investigated [9,10]. While FNNs are known to tend to overfitting due to the huge number of parameters, CNN layers are not optimally suited for processing of time-resolved sequences.

---

[1] Corresponding Authors, Pattern Recognition Lab, Department of Computer Science, Friedrich-Alexander-Universität Erlangen-Nürnberg, Erlangen, Germany; E-mail: elisabeth.hoppe@fau.de, florian.thamm@fau.de.

In this work, we show the applicability of Recurrent Neural Networks (RNNs) for solving the MRF reconstruction problem. The acquired signals are correlated in time, thus, RNNs are well suited for this reconstruction problem. They can provide a more effective structure than simple CNN layers, as they are capable of memorization of temporal structures within the sequences. We evaluate our method using in-vivo data from one healthy volunteer and compare it to another deep learning method based on a CNN [6].

## 2. Methods

*2.1. Data*

Data were acquired as axial brain slices in one healthy volunteer (female, 24 years) on a MAGNETOM Skyra 3T MR scanner (Siemens Healthcare, Erlangen, Germany) using a prototype sequence based on Fast Imaging with Steady State Precession (FISP) with spiral readout and the following parameters: Field-of-view (FOV) 300 mm, resolution $1.17\ x\ 1.17\ x\ 5.0$ mm$^3$, variable TR $(12 - 15$ ms), FA $(5 - 74°)$, number of repetitions: 3000, resulting in fingerprints of length $N = 3000$ [2]. The parameter maps reconstructed with a state-of-the-art method using the uncompressed fingerprints and the dictionary [1,2] were used as the ground truth for our training and testing. We used a fine-resolved dictionary ($T_1$: $50 - 4500$ ms, $T_2$: $20 - 800$ ms, $B_1^+$: $70 - 130$ %, overall about 131,000 fingerprints) in order to provide accurate ground truth data.

*2.2. Deep Learning Models*

In order to implement RNNs for this task, we used a commonly known technique, namely Long Short-Term Memory (LSTM) layers [11]. It is known, that like other recurrent layers, the performance of a LSTM layer suffers from too long sequences [12]. Thus, keeping the sequence length in moderate sizes is recommended. Experiments have shown that by applying the LSTM layer directly on the whole fingerprint sequence (3000 data points as sequence axis) - independent from the subsequent design of the network - the model did not solve the regression problem. As well as (1), the large sequence length of 3000 steps as (2) the small content per time step (one value in case of magnitude, two values in case of complex data) counteracts the total regression performance. Reshaping the sequence into 30 even sized parts, each consisting of 100 time points in alternated order in front of the LSTM layer as the first layer of the network, solves these limitations at once. The reduction from 3000 to 30 iterations considerably decreases the risk of having vanishing/exploding gradient related problems, furthermore improving the total inference and training speed. At the same time, each step contains 100 (amplitude) or 200 (real and imaginary) data points; therefore, reshaping increases the amount of information per sequence element. Afterwards, one single LSTM is applied followed by ReLU and BatchNormalization [13]. Multiple LSTM layers in subsequent order performed disadvantageous in our reshaped setup. Our best perfoming architecture is summarized in Table 1.

**Table 1.** RNN model. The magnitude case is abbreviated with 'M' and the complex case with 'C'. After each computational layer follow ReLU and BatchNormalization layers, but they were omitted for reasons of clarity.

| Layer | Output Shape |
|---|---|
| Input | M: 3000 x 1 |
|  | C: 3000 x 2 |
| Reshape | M: 30 x 100 |
|  | C: 30 x 200 |
| LSTM | 30 x 300 |
| Flatten | 9000 |
| Dense | 2000 |
| Dense | 1333 |
| Dense | 666 |
| Dense | 2 |

*2.3. Training and Evaluation*

In total 8 axial brain slices as complex valued signals were measured as described in Section 2.1, where 6 of them were used for training, one for validation and one for testing. We trained the RNN and the CNN models using magnitude and complex input signal courses each for 100 epochs. We used the mean squared error as the loss function and optimized by ADAM [14] on a Nvidia GeForce GTX 1060 GPU using Python APIs of Keras [15] and TensorFlow [16].

For the evaluation, differences between ground truth and predicted $T_1$ and $T_2$ relaxation times were measured with the absolute mean error $\mu_{abs}$ and the absolute standard deviation $\sigma_{abs}$, where $t_{pred}$ denotes the $T_1$ or $T_2$ prediction and $t_{gt}$ the $T_1$ or $T_2$ ground truth of in total $N$ fingerprint sequences, computed as follows (Eqs. (1) and (2) ):

$$\mu_{abs} = \frac{1}{N} \sum_{i=0}^{N} |t_{i,pred} - t_{i,gt}| \qquad (1)$$

$$\sigma_{abs} = \sqrt{\frac{1}{N} \sum_{i=0}^{N} (|t_{i,pred} - t_{i,gt}| - \mu_{abs})^2} \qquad (2)$$

The best model based on validation loss out of 100 training epochs was used for testing. All different models were tested using one measurement excluded during the training and validation processes. We compared our RNN model to the CNN based reconstruction described in [6] with small architectural changes to further improve the results (BatchNormalization layers after each ReLU, see Table 2).

**Table 2.** Improved CNN Model based on [6]. The magnitude case is abbreviated with 'M' and the complex case with 'C'. After each computational layer follow ReLU and BatchNormalization layers, but they were omitted for reasons of clarity. Every convolution uses 'same' mode for padding. Kernel size is abbreviated with 'KS' and stride with 'S'.

| Layer | Output Shape |
|---|---|
| Input | M: 3000 x 1 |
|  | C: 3000 x 2 |
| Conv1D (KS: 15, S:5) | 598 x 30 |
| Conv1D (KS: 10, S:3) | 197 x 60 |
| Conv1D (KS: 5, S:3) | 97 x 120 |
| Conv1D (KS: 3, S:2) | 48 x 240 |
| Average Pooling | 23 x 240 |
| Flatten | 5520 |
| Dense | 1000 |
| Dense | 500 |
| Dense | 300 |
| Dense | 2 |

## 3. Results

The comparison between the CNN and the RNN model regarding their performance on the test data set is shown in Table 3. The best reconstruction results were achieved using complex data and the RNN model. We were able to reduce the absolute mean error in $T_1$ by 67% and in $T_2$ by 60% compared to the CNN architecture evaluated on magnitude signals [6]. Using the CNN architecture on complex data, the absolute mean error can be reduced in $T_1$ by 60% and in $T_2$ by 50% compared to its magnitude counterpart. Exemplary qualitative in-vivo results of the test data set are shown in Figures 1 and 2.

The inference time for each reconstruction method was measured on a 2.4 GHz Intel Xeon E5620 and a Nvidia GeForce GTX 1070. The regression of one single sequence with the pattern matching [1] took 30.3 ms (CPU), the prediction using the CNN model took 1.106 ms (on GPU 0.226 ms), and the prediction using the RNN model took 0.984 ms on CPU (on GPU 0.296 ms).

**Table 3.** Error on complex and magnitude sequences of the same test data set. The CNN and RNN errors were measured with the mean absolute and mean standard deviation error denoted with $\mu_{abs} \pm \sigma_{abs}$ in ms.

| Architecture | $T_1$ Error [ms] | $T_2$ Error [ms] |
|---|---|---|
| CNN Magnitude | 89 ± 160 | 20 ± 48 |
| RNN Magnitude | 82 ± 159 | 18 ± 46 |
| CNN Complex | 36 ± 108 | 10 ± 33 |
| **RNN Complex** | **29 ± 94** | **8 ± 30** |

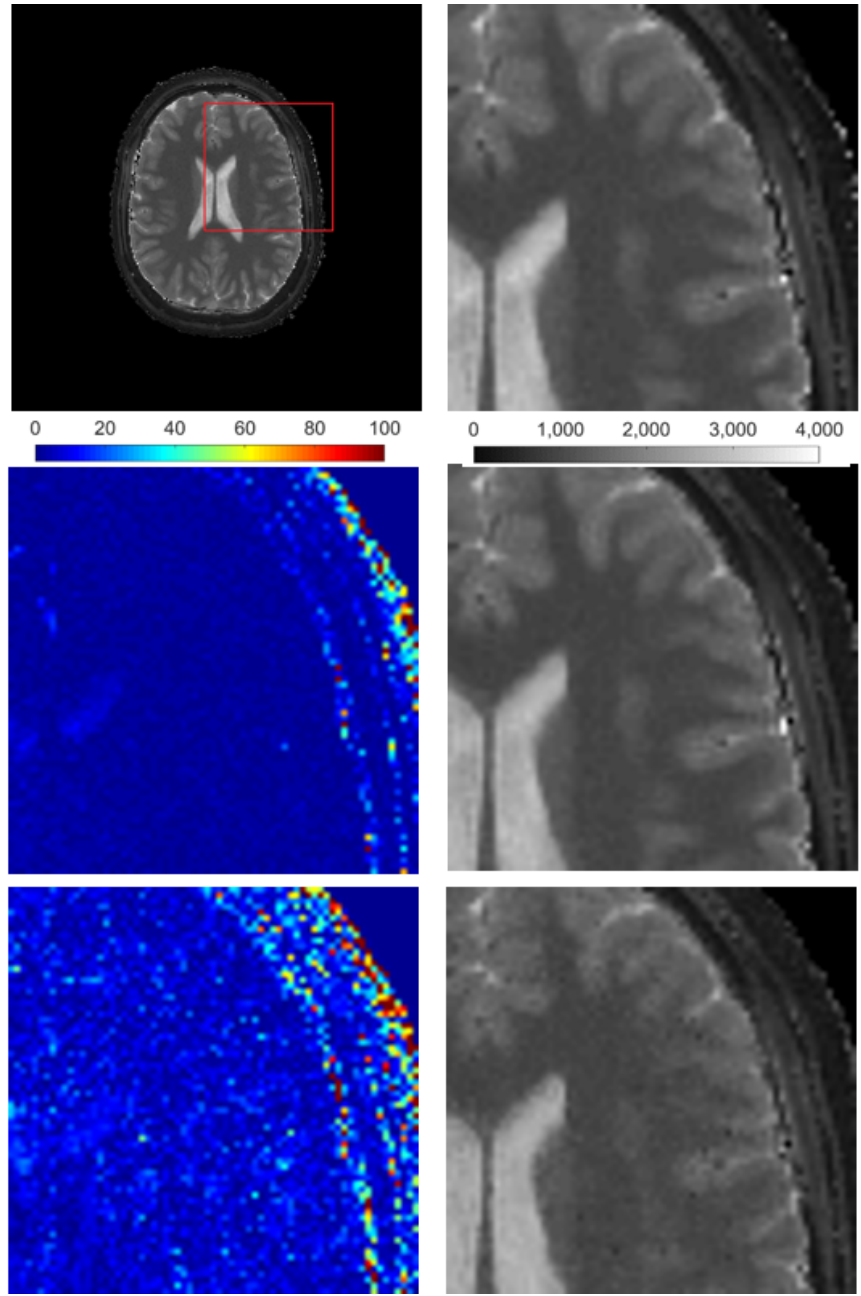

**Figure 1.** Shown are images of the test data set. Top row left: $T_1$ ground truth image. Rectangle marks the region in all other subimages. Top row right: $T_1$ subimage of ground truth. Middle row left: $T_1$ difference between RNN prediction on complex signals and ground truth. Middle row right: RNN predictions on complex signals. Bottom row left: $T_1$ difference between CNN prediction on magnitude signals and ground truth. Bottom row right: CNN predictions on magnitude signals. For better visibility, all relative error maps were clipped at 100 %, the background of all $T_1$ maps was set to -200 and they were windowed equally for fair comparison (0 - 4,000 ms).

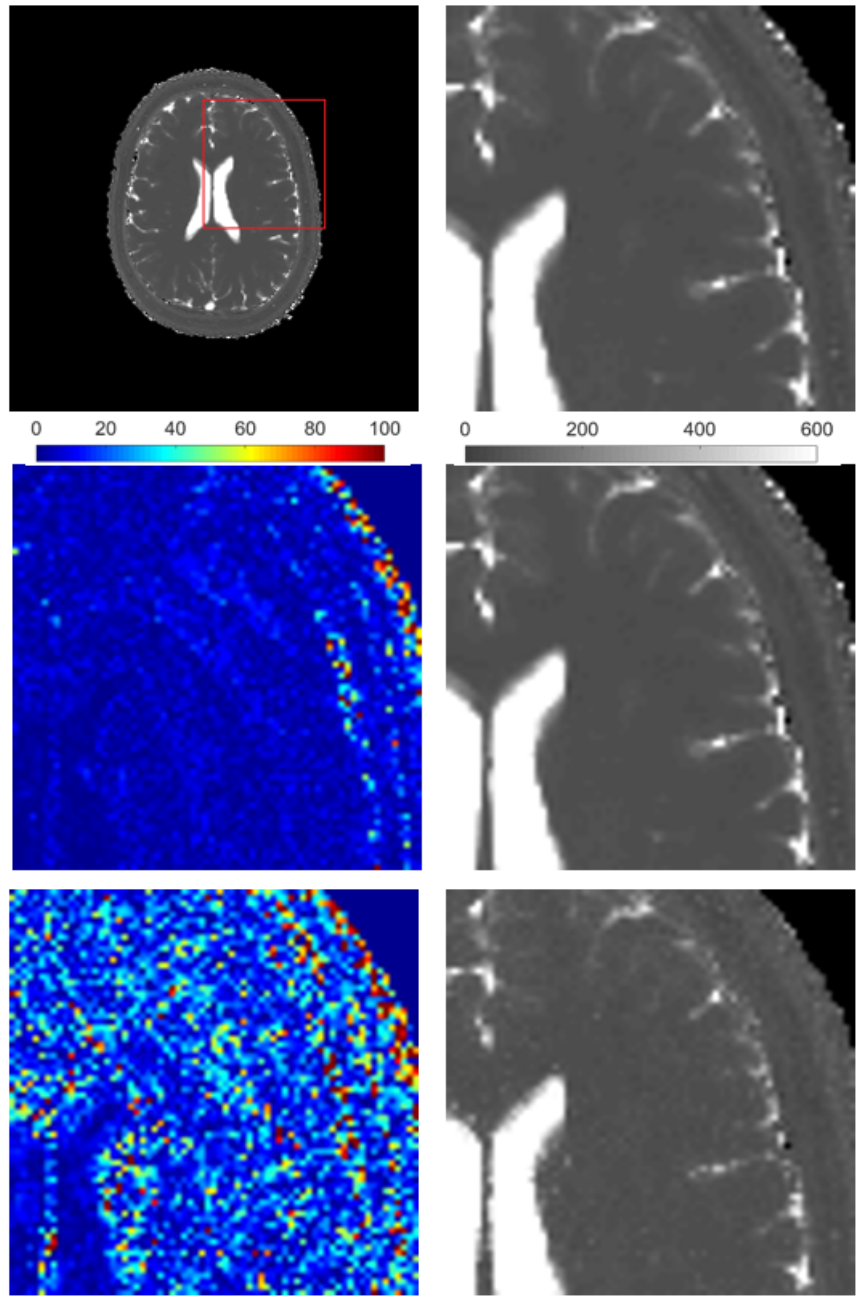

**Figure 2.** Shown are images of the test data set. Top row left: $T_2$ ground truth image. Rectangle marks the region in all other subimages. Top row right: $T_2$ subimage of ground truth. Middle row left: $T_2$ difference between RNN prediction on complex signals and ground truth. Middle row right: RNN predictions on complex signals. Bottom row left: $T_2$ difference between CNN prediction on magnitude data and ground truth. Bottom row right: CNN predictions on magnitude signals. For better visibility, all relative error maps were clipped at 100 %, the background of all $T_2$ maps was set to -200 and they were windowed equally for fair comparison (0 - 600 ms).

## 4. Discussion and Conclusion

Our qualitative results in Figures 1 and 2 show, that in comparison to a CNN, a RNN is able to reduce the prediction errors. Furthermore, the errors mostly are distributed at regions with high noise even in the ground truth maps, especially at borders from soft tissues to the brain skull. In contrary, a CNN predicts a map with errors which are increased and distributed over all tissues. From our quantitative results, two main aspects can be observed: First, DL reconstruction results can be significantly improved by using both components of the complex input signals (like [1]) instead of using only the magnitudes. Although the size of the input layer is increased by a factor of 2, the improvement on the reconstructed images is significant and shows the importance of both components contained in the complex-valued signals. Second, the RNN architecture outperforms previously published CNN architecture, as results obtained from the RNN model are hardly distinguishable from the ground truth data. As the fingerprints are signals correlated in time, RNNs provide a well-suited solution for solving the regression task in MRF reconstruction. Furthermore, the reconstruction based on neural networks is significantly faster than the pattern matching approach – up to a factor of 30. To achieve a faster reconstruction using the approach in [1], the dictionary size would have to be reduced, but this would lead to substantial errors in the parameter maps due to the discretization. For DL approaches, however, the reconstruction time remains the same independent of the different parameter combinations contained in the training data. The burden of long computation time is moved away from the reconstruction itself to the training, but this is done only once.

Although we showed the successful applicability of RNNs and their improvements for MRF reconstruction in this work, following limitations of the current status must be noted: We compared our DL approaches with ground truth provided by the template matching method [1]. While these quantitative maps can be erroneous due to known drawbacks of this template matching method [3], our comparison remains fair, as we have the same ground truth basis for all our methods. Furthermore, the dictionary was simulated using a considerably large amount of entries to reduce such matching errors. Future work will include the comparison of our approach to another quantitative MR techniques or using phatoms with provided quantitative ground truth values. Another drawback is the limited used data here. A future confirmation of our method on a large data set containing brain slices from different volunteers is another important step to show the generalization ability over multiple objects and different anatomies. Furthermore, a systematic evaluation of different kinds and compositions of RNN architectures will be investigated for their influence on the results.

**Conflict of interest:** F. Thamm, C. Syben, F. Schirrmacher and A. Maier have no conflict of interest. E. Hoppe is funded by Siemens Healthcare GmbH, Erlangen, Germany. G. Körzdörfer, M. Nittka, J. Pfeuffer and H. Meyer are employees of Siemens Healthcare GmbH, Erlangen, Germany.

**Informed consent:** Informed consent has been obtained from the individual included in this study.

## References

[1]   Ma, Dan, et al. "Magnetic resonance fingerprinting." *Nature* 495.7440 (2013): 187.

[2] Jiang, Yun, et al. "MR fingerprinting using fast imaging with steady state precession (FISP) with spiral readout." *Magnetic resonance in medicine* 74.6 (2015): 1621-1631.
[3] Wang, Zhe, et al. "MRF denoising with compressed sensing and adaptive filtering." *2014 IEEE 11th International Symposium on Biomedical Imaging (ISBI)*. IEEE, 2014.
[4] Cohen, Ouri, et al. "MR fingerprinting deep reconstruction network (DRONE)." *Magnetic resonance in medicine* 80.3 (2018): 885-894.
[5] Hoppe, Elisabeth, et al. "Deep Learning for Magnetic Resonance Fingerprinting: A New Approach for Predicting Quantitative Parameter Values from Time Series." *GMDS*. 2017.
[6] Hoppe, Elisabeth, et al. "Deep Learning for Magnetic Resonance Fingerprinting: Accelerating the Reconstruction of Quantitative Relaxation Maps." In: *Proceedings of the International Society for Magnetic Resonance in Medicine*. Paris, France; 2018, p.2791.
[7] Fang, Zhenghan, et al. "Deep learning for fast and spatially-constrained tissue quantification from highly-undersampled data in magnetic resonance fingerprinting (MRF)." *International Workshop on Machine Learning in Medical Imaging*. Springer, Cham, 2018.
[8] Song, Pingfan, et al. "HYDRA: Hybrid Deep Magnetic Resonance Fingerprinting." *arXiv preprint arXiv:1902.02882*(2019).
[9] Barbieri, Marco, et al. "Circumventing the Curse of Dimensionality in Magnetic Resonance Fingerprinting through a Deep Learning Approach." *arXiv preprint arXiv:1811.11477*(2018).
[10] Virtue, Patrick, et al. "Better than real: Complex-valued neural nets for MRI fingerprinting." *2017 IEEE International Conference on Image Processing (ICIP)*. IEEE, 2017.
[11] Hochreiter, Sepp, et al. "Long short-term memory." *Neural computation* 9.8 (1997): 1735-1780.
[12] Pascanu, Razvan, et al. "On the difficulty of training recurrent neural networks." *International conference on machine learning*. 2013.
[13] Ioffe, Sergey, et al. "Batch normalization: Accelerating deep network training by reducing internal covariate shift." *arXiv preprint arXiv:1502.03167* (2015).
[14] Kingma, Diederik P., et al. "Adam: A method for stochastic optimization." *arXiv preprint arXiv:1412.6980* (2014).
[15] Chollet, Francois et al. "Keras" https://keras.io, 2015 (accessed: 03/15/2019)
[16] Abadi, Martín, et al. "Tensorflow: A system for large-scale machine learning." *12th {USENIX} Symposium on Operating Systems Design and Implementation ({OSDI} 16)*. 2016.